# Accelerating Discovery of Solid-State Thin-Film Metal Dealloying for 3D Nanoarchitecture Materials Design through Laser Thermal Gradient Treatment

Cheng-Chu Chung[1], Ruipeng Li[2], Gabriel M. Veith[3], Honghu Zhang[2],
Fernando Camino[4], Ming Lu[4], Nikhil Tiwale[4], Sheng Zhang[5],
Kevin Yager[4], Yu-chen Karen Chen-Wiegart[1, 2]

1 Department of Materials Science and Chemical Engineering, Stony Brook University, Stony Brook, NY 11794, USA.
2 National Synchrotron Light Source II, Brookhaven National Laboratory, Upton, NY 11973, USA.
3 Chemical Sciences Division, Oak Ridge National Laboratory, Oak Ridge, TN 37831, USA,
4 Center for Functional Nanomaterials, Brookhaven National Laboratory, Upton, NY 11973, USA.
5 Advanced Science Research Center, The Graduate Center of the City University of New York, New York, NY 10031, USA

*Corresponding Author: Karen.Chen-Wiegart@stonybrook.edu

## Abstract

Thin-film solid-state metal dealloying (thin-film SSMD) is a promising method for fabricating nanostructures with controlled morphology and efficiency, offering advantages over conventional bulk materials processing methods for integration into practical applications. Although machine learning (ML) has facilitated the design of dealloying systems, the selection of key thermal treatment parameters for nanostructure formation remains largely unknown and dependent on experimental trial and error. To overcome this challenge, a workflow enabling high-throughput characterization of thermal treatment parameters while probing local nanostructures of thin-film samples is needed. In this work, a laser-based thermal treatment is demonstrated to create temperature gradients on single thin-film samples of Nb-Al/Sc and Nb-Al/Cu. This continuous thermal space enables observation of dealloying transitions and the resulting nanostructures of interest. Through synchrotron X-ray multimodal and high-throughput characterization, critical transitions and nanostructures can be rapidly captured and subsequently verified using electron microscopy. The key temperatures driving chemical reactions and morphological evolutions are clearly identified within this framework. While the oxidation process may contribute to nanostructure formation during thin-film treatment, the dealloying process at the dealloying front involves interactions solely between the dealloying elements, highlighting the availability and viability of the selected systems. This approach enables efficient exploration of the dealloying process and validation of ML predictions, thereby accelerating the discovery of thin-film SSMD systems with targeted nanostructures.

# 1. Introduction

Dealloying as spontaneous, self-organizing process can be used to fabricate bi-continuous porous or composite structures by selectively removing one component from a parent material into the dealloying agent. These materials offer a range of benefits, such as enhanced mechanical properties, lightweight, high surface area-to-volume ratio, and improved thermal and electrical conductivity.[1-5] Based on the dealloying agent used in the process, dealloying methods can be primarily categorized into electrochemical dealloying,[6-8] liquid metal dealloying,[9-12] solid-state metal dealloying (SSMD),[13-15] molten salt dealloying,[16-18] and vapor phase dealloying.[19-22] Consequently, the dealloyed materials have found applications across a wide range of fields, extensively studied and implemented in energy storage and conversion systems,[1, 23-25] catalysis,[26-30] actuation,[31, 32] biomedical technologies,[33-36] and various sensing devices.[37-39]

Among the various dealloying techniques, thin-film SSMD, an extension of the previously established SSMD process used in bulk materials, has emerged as a promising method for fabricating nanoarchitectured materials.[40-42] In the thin-film SSMD process, a thin-film solvent metal C serves as the dealloying agent. This agent is applied to induce phase separation within a thin-film parent alloy A-B, resulting in the formation of distinct A and B-C components. This novel process can be used to create bi-continuous nanoporous structures and nanocomposites with materials that may not be accessible by other methods. The thin-film geometry of this technique distinguishes itself through several key advantages. First, thin-film SSMD enhances control over the dealloying process by effectively preventing issues with gradient pore size and morphological inhomogeneities that are common in thick bulk materials. Second, using thin-film processes enables the fabrication of nanostructured materials that can be directly integrated into existing thin-film-based technologies. Furthermore, the process can be conducted in a relatively shorter time, facilitating the creation of finer feature sizes.

Despite the promise of thin-film SSMD for use across a wide range of materials and applications, exploring new dealloying systems remains challenging due to the complex interplay of critical parameters that govern morphological, chemical, and structural evolution within a vast parameter space.

To address this challenge, a framework has been proposed that uses machine learning (ML) both to predict potential dealloying systems and to operate autonomous synchrotron X-ray diffraction (XRD) for materials characterization.[43] Although preliminary proof-of-concept results have shown promise, further demonstration of its effectiveness, particularly in experimental validation, remains a challenge. While a subsequent study aimed to one of the potential systems and revealed oxidation as a key factor driving the formation of bi-continuous nanostructures that needs to be incorporated in thin-film SSMD design,[44] the extent of the challenge in thin-film SSMD design becomes more apparent:

Firstly, preparing dealloying thin films under discrete thermal conditions, including various dealloying temperature and time, can be difficult to capture critical dealloying transitions across the thermal parameter space. The process often relies on serendipitous outcomes from multiple dealloying temperature and time settings, without assurance that the key transitions can be identified within these conditions. Furthermore, while synchrotron XRD can effectively provide crystalline phase information, it is unable to capture essential morphological details and local non-crystalline chemical and structural transitions. These aspects are crucial for understanding dealloying processes—morphological data is particularly important because the primary objective of dealloying is to create bi-continuous nanostructures. Additionally, prior research indicates that local transitions may precede the appearance of long-range crystalline phases.[42] Relying solely on XRD may not yield accurate materials design validation or a thorough mechanistic understanding of dealloying, and yet exhaustive characterization of discrete samples to characterize nanostructure and local chemical information, for instance through electron microscopy, is impractical.

To address these challenges, it is essential to develop methods for more effectively exploring this high-dimensional space. The goal is to systematically identify the processing conditions where critical transitions occur, enabling nanostructure formation in thin-film SSMD systems. This includes developing multimodal characterization methods to capture chemical, morphological and structural evolution. Additionally, it requires establishing analysis frameworks to support future autonomous experimentation. Successfully developing these methods is not only crucial for effectively creating and designing new materials through thin-film SSID, but it is also important for a wide range of nano- and functional materials designs that require thermal treatment.

In this work, a novel thin-film SSMD process utilizing laser heating is presented to explore dealloying transitions and efficiently validate ML predictions presented in the prior study.[43] A linear temperature gradient along stripe samples is generated by an infrared laser absorbed by silicon at the one end of the sample, with the opposite end maintained at a lower temperature by a base heater. Through this innovative thermal environment, dealloying process across the continuous thermal space enables the exploration of critical transitions and key processing parameters. Two dealloying systems, Nb-Al/Sc and Nb-Al/Cu were selected to validate ML predictions, with a focus on capturing the dealloying temperature window and investigating potential dealloying systems. The Al-Sc and Al-Cu pairs (B-C components), chosen for their negative mixing enthalpies,[45] were utilized to drive dealloying of the parent alloy, Nb-Al, with Al-Sc exhibiting a more negative enthalpy than Al-Cu. To confirm the occurrence of dealloying in the system, multimodal synchrotron X-ray techniques was conducted to characterize atomic interdiffusion, chemical reactions, and morphological evolution between the B-C components. Specifically, wide-angle X-ray scattering captures the crystalline phase transitions, while small-angle X-ray scattering analyzes the evolution of nanostructures, and X-ray absorption spectroscopy reveals local chemical changes. These techniques are applied to the gradient-heated stripe sample with continuous temperature variation. Once the transition point is identified, complementary methods such as electron microscopy can then be employed to characterize specific temperature points that might otherwise be missed without a gradient temperature condition. As a result, the laser heating process effectively creating a temperature-gradient sample offers significant advantages over traditional lab furnaces and rapid thermal processing (RTP), including a reduction in the number of experiments required compared to conventional isothermal methods. This makes it particularly suitable for discovering novel dealloyed nanostructures in unexplored dealloying systems. Moreover, the laser heating process can be integrated with synchrotron X-ray characterization, enabling *in situ* experiments for real-time observation of the dealloying process. With its ability to explore high-dimensional processing parameter spaces efficiently, the laser heating method sets the stage for the autonomous discovery of new dealloyed materials.

## 2.    Methods and characterization

### 2.1 Dealloyed stripe thin film with gradient heating condition

A sapphire wafer (C-plane (0001), 0.5 mm) was cut into a 40 mm × 10 mm rectangle and adhered onto a 45 mm × 10 mm silicon (Si) substrate using high temperature carbon paste (PELCO®, Ted Pella, Inc). This Si semiconductor substrate with an appropriate band gap can absorb the 980 nm wavelength infrared radiation (IR) laser, heated to a high temperature. Thin film depositions were conducted on the sapphire-silicon integrated stripe substrates through a magnetron sputtering system (Kurt J. Lesker PVD75) at room temperature. The deposition was performed in a chamber evacuated to ≈ 5E-6 torr and then backfilled with Ar gas to 6E-3 torr. Using direct-current (DC) power, $Ar^+$ ions sequentially bombarded high-purity sputtering targets (3-inch diameter, Kurt J. Lesker Company) of Nb/Al (40/60 at.%, 99.95%), Sc (99.9%), and Cu (99.9%). Stripe samples were prepared by depositing parent alloy thin films (40 at.% metal A and 60 at.% metal B) followed by solvent metal C thin films on sapphire substrates.

The thin-film samples from Oak Ridge National Laboratory were prepared using substrates cleaned by sonication in hexane (Fisher - ACS Grade) and isopropanol (Fisher Scientific – Spectroscopy Grade). To create a binary target, pure metal disks were cut into 15°, 30°, 45°, and 60° wedges using electrical discharge machining (EDM) with a 0.05 mm wire. These wedges from different materials were assembled into a composite disk for sputter deposition. The wedges were cleaned by sanding with 1000 grit $Al_2O_3$ sandpaper (McMaster Carr) to remove surface residue, followed by cleaning in hexane and isopropanol solvents. The composition of the disk could be modified by exchanging wedges between materials A and B, as shown in **Figure S 1**. Metal targets were sourced from commercial vendors. Alloy metals were purchased from the Kurt J. Lesker company, including aluminum (99.99%) and niobium (99.95%). The base layer was grown using a scandium target (American Elements - 99.99%).

The base layer was first deposited via DC magnetron sputtering using a pure metal target at a process pressure of 19 mTorr with an argon flow of 20 sccm (99.9995%, Air Liquide), an applied power of 40 W, and a deposition distance of 7 cm. Without breaking vacuum, the alloy layer was grown in the same chamber via DC magnetron sputtering of the wedge target under identical conditions (19 mTorr process pressure, 20 sccm argon flow, 40 W applied power). An aluminum foil baffle was installed between the two sputter guns to prevent cross-contamination. Before deposition, the chamber was evacuated to a base pressure of <1E-6 Torr, and each target was pre-sputtered for 30 minutes to remove surface oxides.

A photothermal annealer (PTA) instrument was used to subject the stripe samples with a temperature gradient. The laser-based instrument was previously developed by a team at the Center for Functional Nanomaterials (CFN), Brookhaven National Laboratory (BNL), and is located at the at the National Synchrotron Light Source II (NSLS-II) of BNL. While the laser annealing method has been successfully utilized in block copolymer thin films, enabling precise spatial control of the self-assembly process through steep temperature gradients,[46-48] our current work extends its application beyond polymer processing (100-300 °C) to metallic thin film thermal processing (100-800 °C). The PTA is equipped with a clamp as a base temperature control. The base heater and the laser heating zone can thus define a range of temperature gradient. The temperature gradient calibration was conducted by salt thermometrics, as described in **Figure S 2**. The thin film of the as-deposited stripe was then fixed to the PTA clamp at one edge and heated by the IR laser at the opposite end. The calibrated temperature gradient from 100.0 to 817.0 °C was held for 30 min to induce dealloying. A 30-minute treatment duration was experimentally verified to be adequate to produce both chemical and morphological alterations in the thin-film structure.[41] The heating chamber was filled with a forming gas (99.999% Airgas, 2.92 % $H_2$, balanced with Ar) prior to the heating, with the gas continuous flowing during the heating process to minimize the oxidation of the thin films. At the end of the thermal treatment, the laser and the base heater were turned off and the sample was immediately cooled down to room temperature in the forming gas environment, with a cooling fan.

## 2.2 High-throughput phase and morphological characterization via grazing-incidence wide-/small-angle X-ray scattering

The thin-film stripe samples were characterized by synchrotron X-ray scattering techniques in a grazing-incidence geometry from 0.1-0.5 degrees at Complex Materials Scattering (CMS, 11-BM), at the NSLS-II, BNL. The samples were mounted onto a bar holder, which was placed on a stage aligned with the X-ray beam. The sample stage was in a vacuum chamber below 1E-3 torr to minimize scattering from the air. The measurement was carried out by a 15.0 keV X-ray beam (~ 0.8266 Å in wavelength) with a 1s exposure time. Each scan was performed along the temperature gradient using a 0.4 mm step size along the length of the stripe sample, to probe the dealloying area within the smallest possible temperature window. The scattering patterns were captured with a Dectris Pilatus 800k and 2M detectors with a pixel size of 172 μm and 0.26 m and 5 m sample-to-detector distance for grazing-incidence wide-/small-angle X-ray scattering (GIWAXS/GISAXS), respectively. A 1D profile data was obtained by integrating a 2D circular pattern from 0.004 – 0.1 Å and 0.7 – 4.2 Å in small- and wide-angle range, respectively. The sample-to-detector distance was calibrated via a measurement from standard cerium oxide power (NIST standard, SRM 674b).

## 2.3 Chemical states and atomic coordinates analysis through X-ray absorption fine structure

The chemical status and bonding information of the thin-film materials were analyzed using X-ray absorption near-edge structure (XANES) spectroscopy, at the Beamline for Materials Measurement (BMM, 6-BM) at the NSLS-II, BNL. The thin film was mounted laterally on a vertical stage, with the thermal gradient direction perpendicular to the beam path. A grazing-incidence geometry was utilized to enhance surface sensitivity for analyzing the thin films. The XANES spectra was collected at the Nb, Sc, and Cu absorption edges using fluorescence mode. The measurements focused on specific locations where diffraction patterns indicated the formation of new phases in the samples based on the analysis from the GIWAXS/GISAXS measurements. Prior to each sample scan, spectroscopic references including Nb, Cu foils and Sc powders (99.9%, Sigma-Aldrich) were measured in transmission mode for energy calibration and subsequent data processing. To enhance the signal-to-noise ration of the absorption spectra, multiple scans were collected for each sample location and averaged. The number of scans depended on the total counts in fluorescence mode: two scans for the Nb K-edge, two for Cu K-edge, and six for Sc K-edge.

## 2.4 Surface and cross-sectional characterization by electron microscopy

Cross-sectional morphology of the thin-film samples was examined using a dual-beam scanning electron/focused ion beam microscope (FIB-SEM FEI Helios), at the Center for Functional Nanomaterials (CFN) at BNL. The instrument was also used to prepare lift-out specimens for transmission electron microscopy (TEM) following standard procedure.

Nanoscale characterization of the thin film was performed using a high-resolution analytical scanning transmission electron microscope (STEM) (FEI Titan Themis 200 kV TEM) operated at 200 kV, at the Advanced Science Research Center at the City University of New York (CUNY) Graduate Center. The STEM, equipped with high-angle annular dark-field imaging (HAADF) and energy-dispersive X-ray spectroscopy (EDS), was employed to analyze both pristine and dealloyed samples. This analysis allowed for imaging of thin-film features and mapping of elemental distribution resulting from the dealloying process.

## 3.      Results and discussion

### 3.1 Design of dealloyed thin-film stripe sample with gradient heating condition.

The dealloying process of the Nb-Al alloy thin film was induced by three different dealloying agents: α-Sc, β-Sc, and Cu solvent metal. They were chosen to study the effect of the phase and elemental composition of the dealloying agent on the dealloying processes, and to systematically validate and improve the ML predictions. The rationale for this design is described below. A ML-predicted Nb-Al/Sc dealloying system was found to undergo dealloying via an oxidation process and form a bi-continuous nanocomposite at  high temperature.[44] In the prior study, the β-Sc phase was identified in the pristine Sc metal used as the dealloying agent. Therefore, a different starting phase of the solvent metal, α-Sc, was synthesized for use as a dealloying agent. Comparing the dealloying processes induced by α-Sc and β-Sc helps to understand the impact of the initial thin-film structure on the phase evolution and subsequent nano-architecture morphology. Moreover, the prior study showed limited interaction between the solvent metal β-Sc and the parent alloy, leading to the introduction of another dealloying agent Cu, in the current study. Nb-Al/Cu was selected because it was also predicted by the ML models as a potential system in the prior work,[43] albeit with a less negative mixing enthalpy for the Al-Cu (-1 kJ/mol), compared to the Al-Sc pair (-38 kJ/mol).

To explore the dealloying process and dealloyed structures across a wide range of thermal treatment, a workflow of accelerating the discovery of the SSMD materials is shown in **Figure 1**. The creation of dealloyed stripe samples with a gradient heating condition covering from 100.0 to 817.0 °C was prepared by a thin-film deposition and followed by an IR laser heating. This gradient heat treatment enables the investigation of the dealloying transition in a continuous thermal space. A high-throughput characterization is then performed using multimodal synchrotron X-ray techniques, enabling identification of phase transition and nanostructure formation at specific temperatures along the temperature gradient. Most importantly, it captures subtle structural changes within a narrow reaction temperature range that can be challenging to discover if only discrete samples were tested. Additionally, at the higher temperature, this method can probe the threshold temperature where the oxidation process begins. Finally, the key transition temperatures identified in the multimodal X-ray analysis can then be further validated and characterized in detail using complementary techniques such as electron microscopy. The results from this detailed characterization can then be fed back into the ML model to refine and improve the selection process for subsequent materials design cycle. More detailed information regarding the fabrication process is described in the **Methods and Characterization Section**.

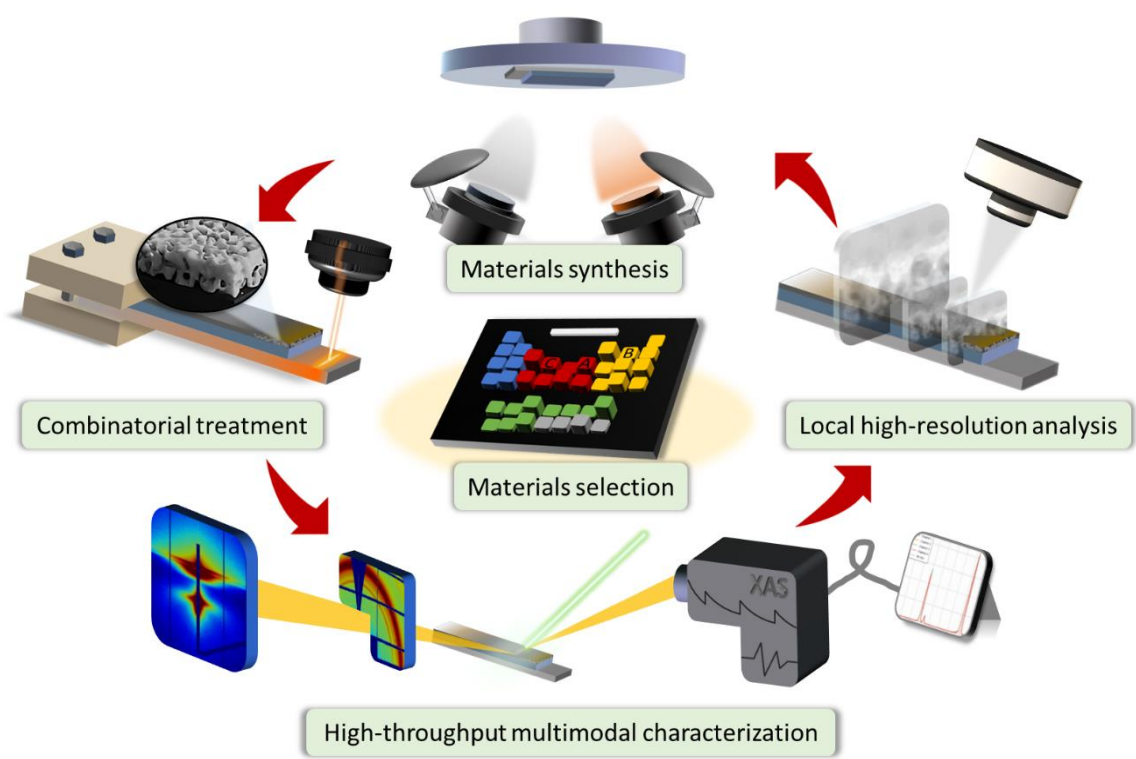

**Figure 1.** Workflow for accelerating the discovery of nanostructured materials using thin-film SSMD materials as a proof-of-concept demonstration. The process begins with ML-guided selection of potential systems (dealloying systems for SSMD). These selected systems are used for materials synthesis (thin-film SSMD samples) under controlled combinatorial treatment (temperature-gradient conditions here) to create a bi-continuous nanostructure (shown in zoom-in inset). Using high-throughput multimodal characterization (multiple synchrotron X-ray techniques here), the samples are characterized to capture nanostructure formation at specific temperatures and further verified using local high-resolution analysis (STEM imaging here). The experimental results verified through electron microscopes are subsequently fed back into the ML model, designing the next potential systems.

## 3.2 Characterization of the phase evolution along the temperature gradient

Owing to the power of the high-throughput characterization conducted by grazing-incidence wide-angle X-ray scattering, GIWAXS, the phase evolution of the stripe thin-film samples can be efficiently measured as a function of dealloying temperature, as shown in **Figure 2** and **Figure S 3**. In the pristine thin films, a broad parent alloy peak and several peaks corresponding to their respective solvent metals, α-Sc, β-Sc, and Cu, represent the starting phases. As the temperature rises along the stripe, each set of thin film sample undergoes a clear phase evolution and new phase formation. Note that although forming gas was used to create a reducing environment, trace amounts of oxygen were still sufficient to promote oxidation at the higher temperature range.

In the Nb-Al/α-Sc system (**Figure 2a and b**), the solvent Sc transitions from its α-phase through a high-temperature β-phase before ultimately forming $Sc_2O_3$. At the intermediate temperature range around 600 °C, α-Sc transforms into β-Sc and reacts with Al to form the intermetallic compound $AlSc_2$. In the β-Sc system (**Figure 2c and d**), β-Sc combines with Al to form AlSc at approximately 650 °C, while oxidizes to $Sc_2O_3$ at higher temperatures. This finding indicates that a potential dealloying process may happen at a middle-temperature window. Above 700 °C, the entire thin film was dominated by the $NbAl_3$ and $Sc_2O_3$.

As shown in **Figure 2e and f**, while the Cu serves as a solvent in the system, the peak corresponding to Cu disappears with increased temperature, revealing a dissolution of the metal Cu into the Nb-Al parent alloy. Meanwhile, the broad peak corresponding to the parent alloy Nb-Al (marked as $Nb_xAl_y$ in the figure to indicate the changing composition) gradually narrows down to a sharp $Nb_3Al$ phase starting from 600 °C. Above 700 °C, the pure Nb phase starts to form and remains stable in the thin film above this temperature. The formation of the $Nb_3Al$ phase, instead of $NbAl_3$ phases found in Nb-Al/Sc system, suggests that less Al remains in the parent Nb-Al alloy, as more Al preferentially reacts with Cu at higher temperatures.

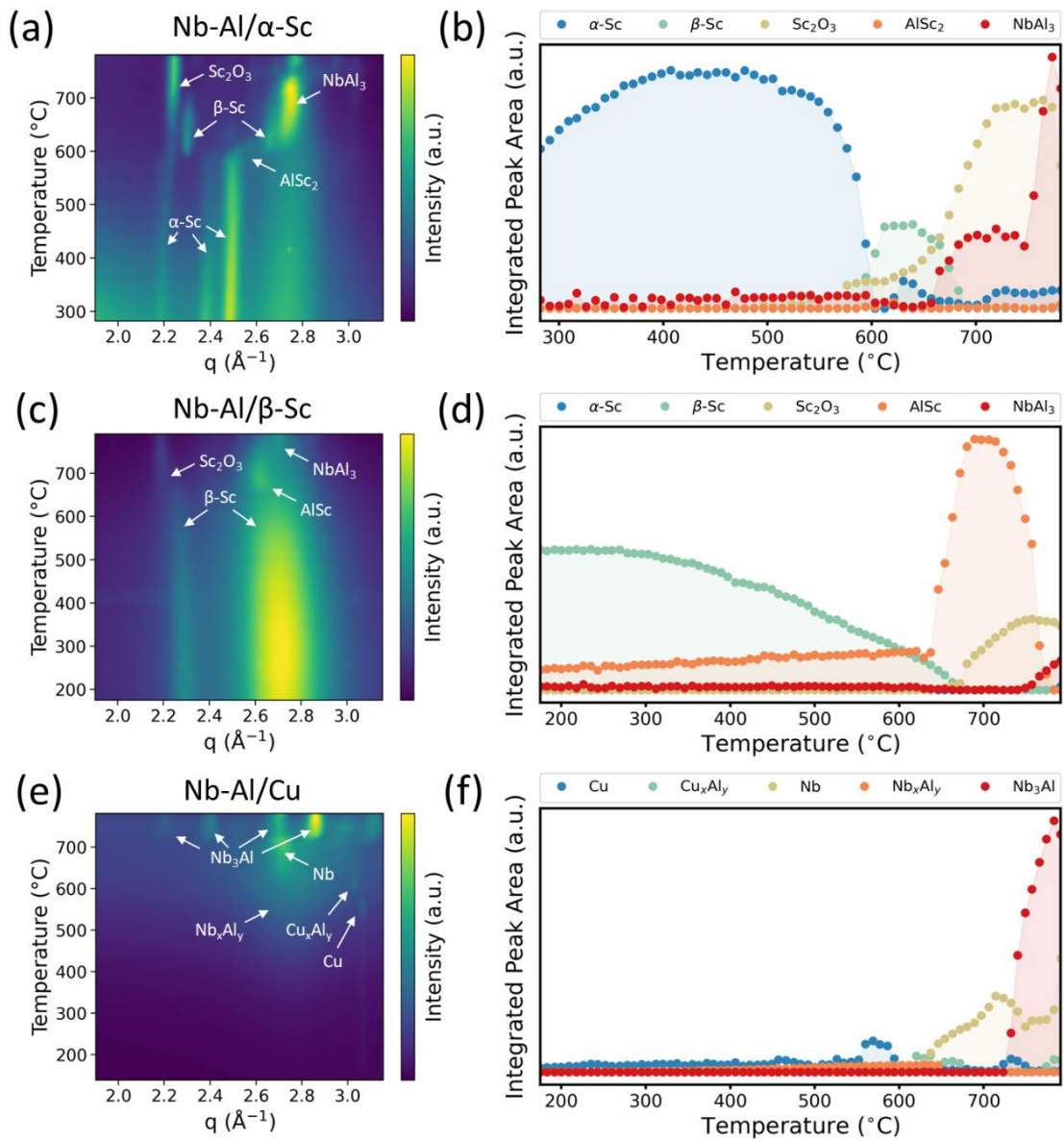

**Figure 2.** Evolution of the diffraction 2D map (a, c, e) and peak area (b, d, f) across stripe samples of Nb-Al/α-Sc, Nb-Al/β-Sc and Nb-Al/Cu thin films, respectively, from low to high temperatures via grazing-incidence wide- (GIWAXS) X-ray scattering. Peak area profiles (b, d, f), corresponding to different phases, are derived from the q-ranges within the bright regions indicated by white arrows in the diffraction 2D map (a, c, e).

### 3.3 Chemical state changes at the critical dealloying transitions.

To gain a deeper understanding of the elemental reactions in the system, the chemical evolution of the reacted thin film at different thermal conditions was studied. X-ray absorption spectroscopy (XAS) was employed to investigate the evolution of the chemical composition and oxidation state of the thin film, as shown in **Figure 3**. The XANES spectroscopic analysis reveals that Nb maintains its pristine state in both the Nb-Al/$\alpha$-Sc and Nb-Al/$\beta$-Sc systems, as shown in **Figure 3a and b**, with the only exception being Nb at approximately 665 °C in the Nb-Al/$\alpha$-Sc system, as indicated by black arrows. The lack of change in the Nb XANES spectra indicate that Nb remains in the alloys with Al without oxidation or dealloying by Sc, while the altered Nb features suggest changes in its coordination environment. In contrast, when the dealloying process is driven by the Cu, the Nb shows a slightly different evolution in **Figure 3c**. Although the Nb resembles the pristine status at low temperatures, it gradually changes to pure Nb above 700 °C, consistent with the GIWAXS results. Regarding the changes in the dealloying agents, as shown in **Figure 3d and e**, both $\alpha$-Sc and $\beta$-Sc solvent metals progressively transform into $Sc_2O_3$ as the temperature increases, as marked by black arrows. Interestingly, in **Figure 3f**, the Cu feature alters significantly with the temperature. Rather than shifting toward an oxidation state, the Cu absorption peak diminishes in intensity, as denoted by black arrows, revealing changes in the local coordination environment for the Cu atoms.

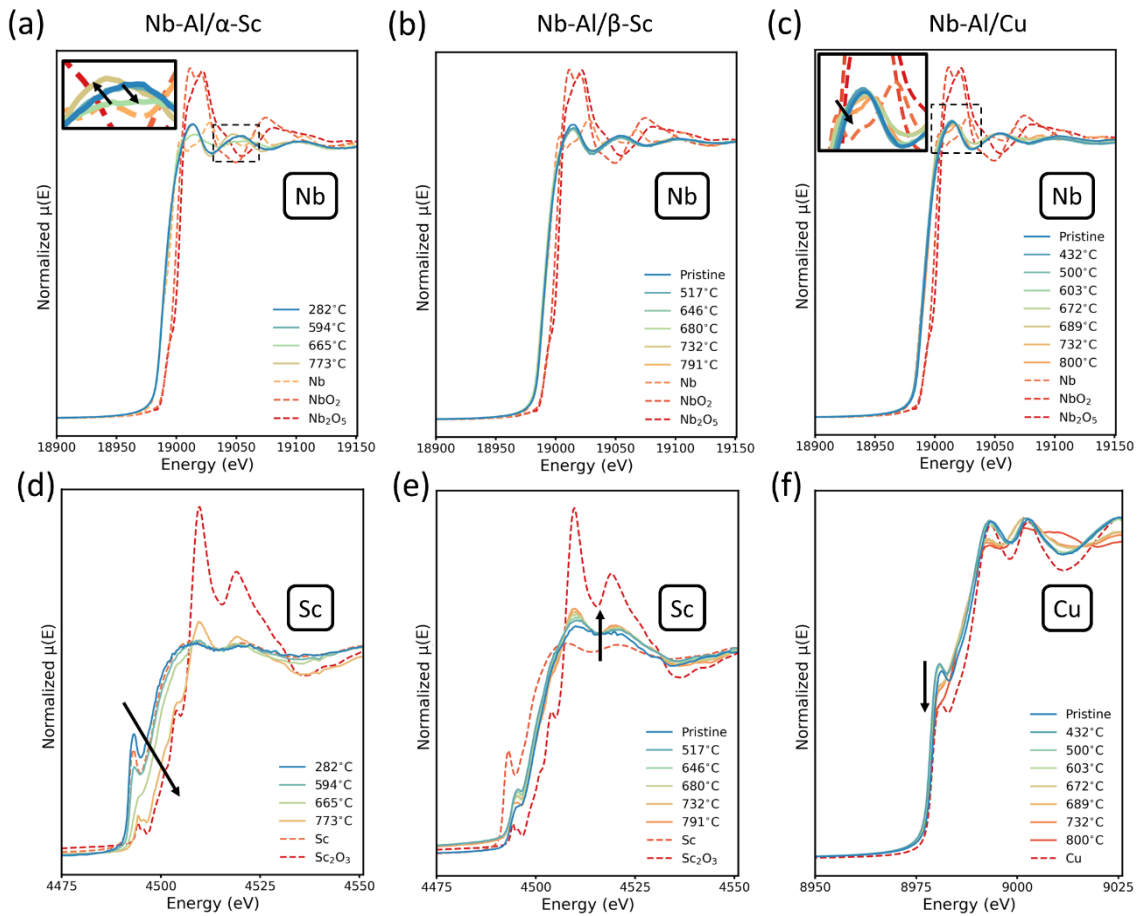

**Figure 3.** Chemical evolution of Nb-Al/α-Sc, Nb-Al/β-Sc, and Nb-Al/Cu thin films regarding the (a-c) remaining element Nb, and (d, e) solvent metals Sc, and (f) Cu K-edge absorption energy at various temperatures, measured under a grazing-incidence geometry. The dashed lines denote standards for various oxidation states, while the black arrows indicate changes with rising temperature. Insets in (a) and (c) show enlarged views of the sections of the plots as indicated for clarity.

### 3.4 Identification of nanostructured morphology and elemental distribution

The high-throughput grazing-incidence small-angle X-ray scattering (GISAXS) results, presented in **Figure 4**, demonstrate a positive correlation between temperature increase and characteristic length growth in the thin film, as evidenced by the shifting position of the main scattering peak. The relationship between the scattering vector $q$ and ligament–ligament spacing $d$ is described by the approximation $d \approx 2\pi/q$, with $d$ roughly corresponding to twice the ligament size due to the similar feature sizes of the two phases in the bi-continuous nanocomposite.[49-52] As temperature increases, the characteristic length shifts toward the lower $q$ direction, corresponding to a larger feature size, as also shown in **Figure S 4**. At approximately 700°C, samples with different compositions exhibit distinct variations in feature sizes. While all systems can achieve characteristic features averaging 10 nm (corresponding to q ≈ 0.03 Å$^{-1}$), the Nb-Al/Cu system exhibits rapid growth with increasing temperature, indicating reactive chemical reactions and atomic diffusion.

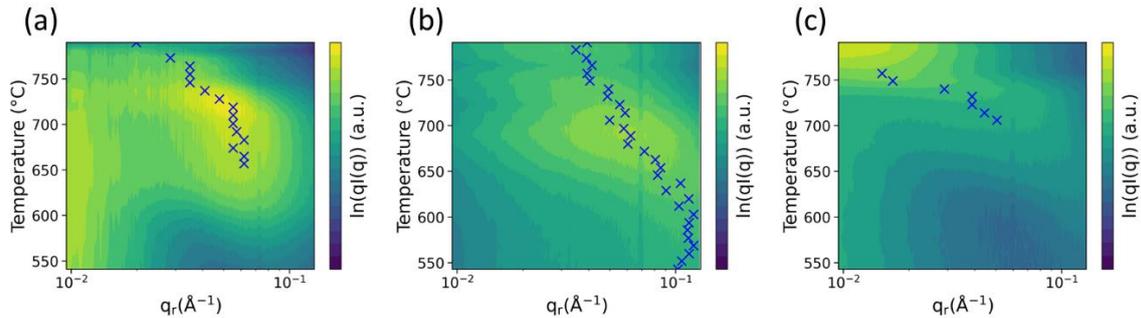

**Figure 4.** Characteristic length maps across thermal space for stripe samples of (a) Nb-Al/α-Sc, (b) Nb-Al/β-Sc, and (c) Nb-Al/Cu thin films, measured by grazing-incidence small-angle X-ray scattering (GISAXS) at an incident angle of 0.25 degrees. Cross markers represent diffusive scattering peaks.

Based on the critical transition points identified in GIWAXS and GISAXS, several positions along the stripe samples were selected, corresponding different dealloying temperatures, to further visualize the nanostructure morphology in cross-section under FIB-SEM. As shown in **Figure 6**, a bi-continuous nanostructure developed within the top portion of the parent alloy. The ligament sizes also match the quantification from the GISAXS analysis, increasing with the rising temperature. The evident contrast of the secondary electron in the parent alloy indicates the phase separation after the thermal treatment.

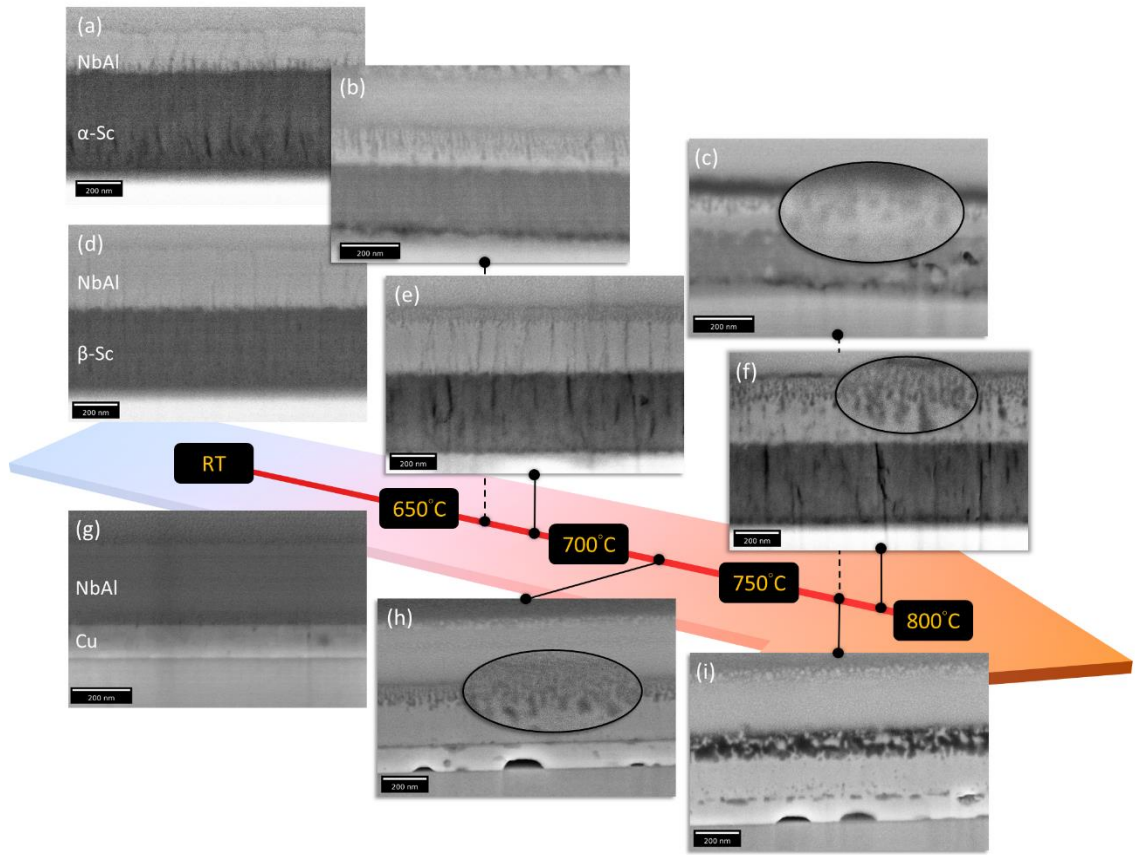

**Figure 5.** Cross-sectional images of pristine (a, d, g) and dealloyed thin films heated at (b) 665°C, (c) 773°C, (e) 689°C, (f) 791°C, (h) 732°C, and (i) 774°C, for 30 minutes with white scale bars at 200 nm. The systems compared include (a-c) Nb-Al/α-Sc, (d-f) Nb-Al/β-Sc, and (g-i) Nb-Al/Cu, examining different starting phases and solvent metal elements.

To further confirm the phase information, an elemental mapping under the STEM with HAADF EDX was conducted to study the elemental distribution as shown in **Figure 6**. When using Sc as the dealloying agents, while Nb and Al elements undergo separation within the parent alloy, Sc remains as a separate layer of thin film, and does not diffuse further into the parent alloy matrix. This behavior is clearly seen in the line scan profiles shown in Figure 6(b) for α-Sc and Figure 6(e) for β-Sc, where the Sc content is minimal in the parent alloy. Meanwhile, although no crystalline phase of $Al_2O_3$ is detected in the diffraction data, the oxygen distribution coincides with the Al distribution (**Figure S 5**), indicating the presence of $Al_2O_3$, possibly in an amorphous form.[53] Additionally, a small amount of Al above the sapphire substrate reveals a minor reaction at the Sc and substrate interface. However, this interfacial reaction has minimal impact on the dealloying process since it occurs far from the active dealloying front. In contrast, the Nb-Al/Cu system in **Figure 6**(g) and (h) shows elemental interdiffusion between Al and Cu. Al diffuses into the entire Cu solvent, while Cu partially diffuses into the parent alloy. This process aligns with the GIWAXS results, where the Cu (111) peak shifts toward a lower angle with increasing temperature, possibly corresponding to Al incorporation into the Cu matrix. The Cu absorption features in the XANES spectrum also indicate local changes in pure Cu-Cu bonding, supporting the observed diffusion process. Despite the absence of nanostructures at the dealloying front, the top surface (**Figure 6**(i)) exhibits phase separation with Al oxidation in the parent alloy.

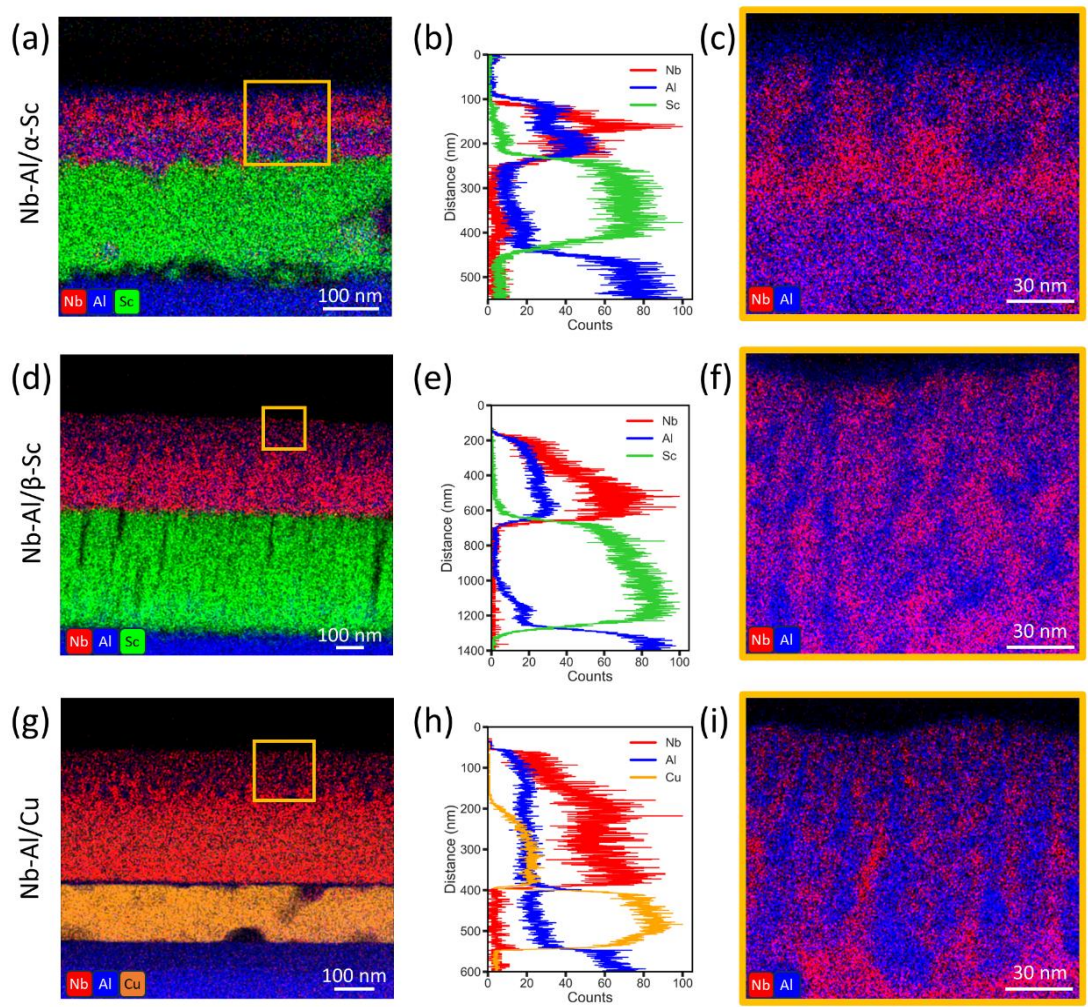

**Figure 6.** STEM characterization using HAADF EDX for the Nb, Al, and Sc elements in (a-c) Nb-Al/α-Sc, (d-f) Nb-Al/β-Sc, and Nb-Al/Cu thin films at 773°C, 791°C, and 732°C, respectively. The figures represent the following: (a, d, g) Elemental distribution of Nb, Al, and Sc/Cu within the thin films. (b, e, h) Line scans from (a, d, g), respectively, from top to bottom. (c, f, i) Zoomed-in views of the top sections of the thin films.

## 3.5 Formation of the nanostructure and dealloying process

The nanostructure forming on the top portion of the thin film results from oxygen diffusion, which matches the overlapping signal of Al and O in the EDX analysis. The feature size grows with the gradient temperature, reaching ~10 nm at a high temperature. In Nb-Al/Sc system, the rest of the region was occupied by the parent alloy and $NbAl_3$ phase. Conversely, instead of diffusing into the parent alloy, the Sc forms an intermetallic compound with Al at the bottom of the layer. This finding suggests a minor interaction happens at the interface between thin film and the $Al_2O_3$ substrate, but it does not affect the dealloying region. In contrast, although the nanostructure resembles that in the Nb-Al/Sc system, the reaction in the Nb-Al/Cu system evolves differently. The Cu has an interdiffusion with Al as the temperature rises according to the peak shifts. The reduced Cu feature in XANES also reflects the change in the local coordination of the Cu near Al. Above 700 °C, the Cu diffraction peak even disappears, leaving the $Nb_3Al$ in the thin film. $Nb_3Al$ also indicates that less Al reacts with Nb, but the Al is dissolved, diffuses into the Cu, or combines with oxygen. These changes reveal that Cu drives the diffusion process while Nb becomes pure Nb metal compared to the Nb-Al/Sc system.

## 4.    Conclusion

In this study, the efficacy of thin-film solid-state metal dealloying (SSMD) processes using infrared (IR) laser-based heating, specifically photothermal annealer (PTA), was demonstrated. A wide range of thermal conditions was explored using a continuous temperature gradient, enabling the investigation of dealloying transitions and morphological evolution in thin-film systems.

Dealloying systems consisting of Nb-Al parent alloy thin films designed for dealloying by Sc and Cu were examined. These systems were subjected to PTA treatment to validate prior machine learning (ML) predictions and to study the underlying mechanisms in thin-film solid-state configurations under key conditions. Complex interactions between thermodynamics, kinetics, and oxidation processes in thin-film SSMD were revealed.

Despite the negative mixing enthalpy of Al-Sc and Al-Cu systems, it was observed that interdiffusion between Al and Cu only initiates at intermediate temperatures. At higher temperatures, while Al-Cu continued to exhibit interdiffusion, it did not form distinct separated phases and the anticipated nanostructure. Instead, an oxidation process was discovered to initiate at the top portion of the parent alloy, leading to interactions between Al and oxygen. This interaction resulted in the formation of a bi-continuous nanostructure with Nb, rather than with the intended solvent metals (Sc or Cu).

Significant growth in the nanostructure feature size along the temperature gradient was observed. This observation indicated the existence of a critical temperature range where key morphological changes occur, highlighting the importance of precise thermal control in SSMD processes.

Contrary to initial expectations, it was found that the Nb-Al parent alloy systems with Sc and Cu as solvent metals did not form the anticipated bi-continuous nanostructure through conventional dealloying mechanisms, despite some interdiffusion between Al and Cu. Instead, trace amounts of oxygen possibly in the reduced environment still promote oxidation on the top part of the parent alloy thin film at the higher temperatures. This oxidation process emerges as a crucial factor in nanostructure formation, resulting in a composite structure of Nb-Al intermetallics and $Al_2O_3$.

These findings contribute significantly to the understanding of thin-film SSMD processes and highlight the importance of considering oxidation effects in high-temperature dealloying, particularly for metals with high oxygen affinity. It is suggested that future work can leverage this oxidation-driven nanostructure formation for potential applications such as nanostructure design of oxide thin films, and exploring strategies to mitigate or control oxidation in SSMD processes where it is undesirable.

Furthermore, this study demonstrates the power of combining high-throughput experimental techniques like synchrotron X-ray characterization, with temperature-gradient treatment like PTA and with machine learning predictions to rapidly explore and understand complex materials systems. With the flexibility of the portable laser instrument, an *in situ* heating treatment can be conducted to observe the real-time dealloying process via synchrotron X-ray multimodal characterization with the aid of autonomous experimentation. This approach is considered promising for accelerating materials discovery and optimization in various fields beyond dealloying, potentially revolutionizing how materials research and development are approached.

**Conflicts of interest**

The authors declare no conflict of interest.

**Acknowledgements**

This work was supported by the National Science Foundation under Grant No. DMR-1752839. The authors acknowledge the support provided via the Faculty Early Career Development Program (CAREER) and the Metals and Metallic Nanostructures Program of the National Science Foundation. A portion of this work (GMV - film growth) was supported by the US Department of Energy's Energy Efficiency and Renewable Energy program, Vehicle Technologies Office, as part of the US-Germany Consortium. This research used resources, Beamline for Materials Measurement (BMM, 6-BM) and Complex Materials Scattering Beamline (CMS, 11-BM) of the National Synchrotron Light Source II (NSLS-II), a U.S. Department of Energy (DOE) Office of Science User Facility operated for the DOE Office of Science by Brookhaven National Laboratory (BNL) under Contract No. DE-SC0012704. This research used Electron Microscopy, Nanofabrication, and Materials Synthesis and Characterization Facilities of the Center for Functional Nanomaterials (CFN), which is a U.S. DOE Office of Science Facility, at Brookhaven National Laboratory under Contract No. DE-SC0012704. The authors are grateful to Dr. Bruce Ravel (National Institute of Standards and Technology), Lead Beamline Scientist at the BMM beamline, for his expertise and support on XAS characterization as well as his insights into data analysis and scientific interpretation. Cheng-Chu Chung acknowledges the support of student fellowship from Joint Photon Sciences Institute (JPSI).

**Author contributions**

C.-C.C., Y.-c.K.C.-W., and K.Y. conceptualized the research. C.-C.C. wrote user proposals for the CMS and BMM beamtime at NSLS-II, BNL, and instruments at CFN with inputs from R. L., K.Y., and Y.-c.K.C.-W. C.-C.C. performed DC sputtering deposition and PTA heating process. K.Y. provided training and supervision for the PTA system. G.V. prepared Nb-Al/α-Sc thin-film materials from ORNL. C.-C.C conducted synchrotron experiments at both CMS and BMM beamlines. R.L., H.Z, and K.Y. provided scientific insights and technical support, as well as guidance on the data analysis, for the CMS experiments. N.T. and M.L. trained C.-C.C. in sputtering deposition operations. F.C. instructed C.-C.C. in cross-sectional imaging characterization and TEM sample preparation using FIB-SEM at CFN. C.-C.C. and Y.-c.K.C.-W. drafted the manuscript, incorporating contributions from all co-authors.

**Accelerating Discovery of Solid-State Thin-Film Metal Dealloying for 3D Nanoarchitecture**

**Materials Design through Laser Thermal Gradient Treatment**

Cheng-Chu Chung[1], Ruipeng Li[2], Gabriel M. Veith[3], Honghu Zhang[2],
Fernando Camino[4], Ming Lu[4], Nikhil Tiwale[4], Sheng Zhang[5],
Kevin Yager[4], Yu-chen Karen Chen-Wiegart[1, 2]

1 Department of Materials Science and Chemical Engineering, Stony Brook University, Stony Brook, NY 11794, USA.
2 National Synchrotron Light Source II, Brookhaven National Laboratory, Upton, NY 11973, USA.
3 Chemical Sciences Division, Oak Ridge National Laboratory, Oak Ridge, TN 37831, USA,
4 Center for Functional Nanomaterials, Brookhaven National Laboratory, Upton, NY 11973, USA.
5 Advanced Science Research Center, The Graduate Center of the City University of New York, New York, NY 10031, USA

*Corresponding Author: Karen.Chen-Wiegart@stonybrook.edu

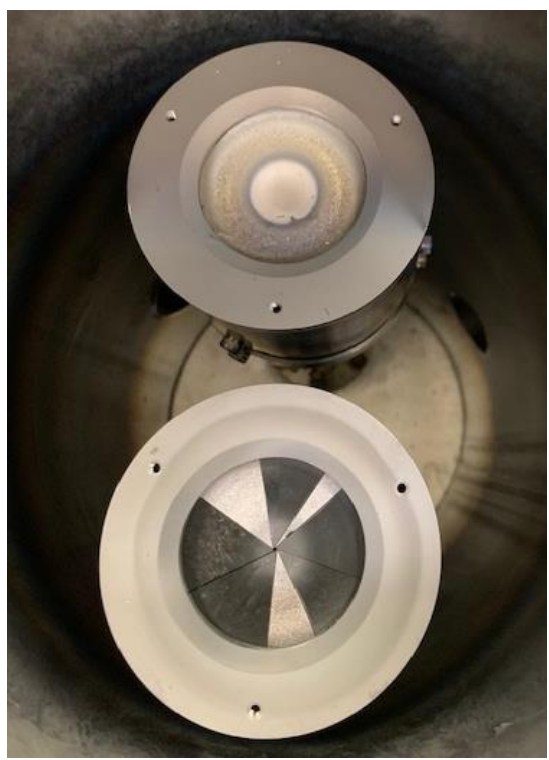

**Figure S 1.** Top-view image of the deposition chamber at Oak Ridge National Laboratory. The bottom sample shows the wedge-shaped target combining Nb (shiny surface) and Al (dull surface), while the top target contains scandium.

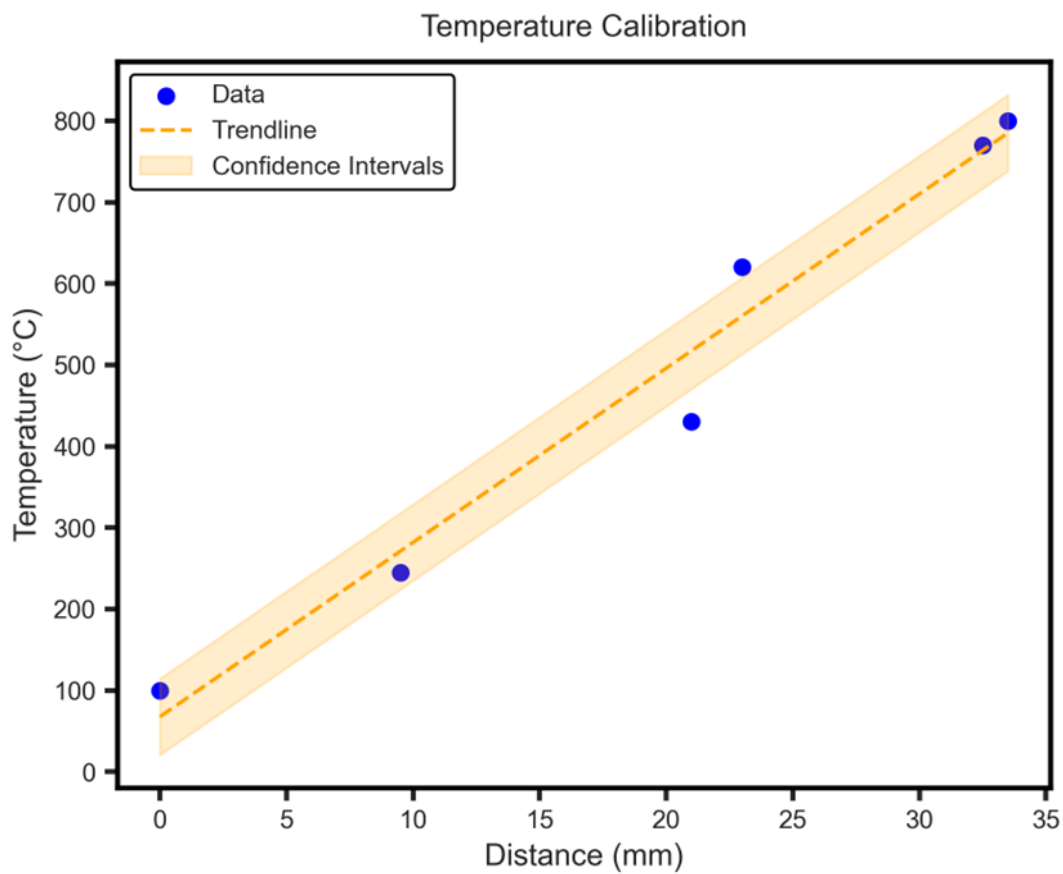

**Figure S 2.** Temperature calibration using salt thermometrics. The temperature profile was determined by identifying the locations where various salts melt: potassium chloride (KCl, 801°C), sodium chloride (NaCl, 770°C), cupric chloride (CuCl2, 620°C), copper(I) chloride (CuCl, 420°C), tetrabutylammonium hexafluorophosphate (TBAHFP, ((CH3CH2CH2CH2)4N(PF6), 245°C), and the base heater temperature (100°C), which is the reference point at the edge of the clamping location.

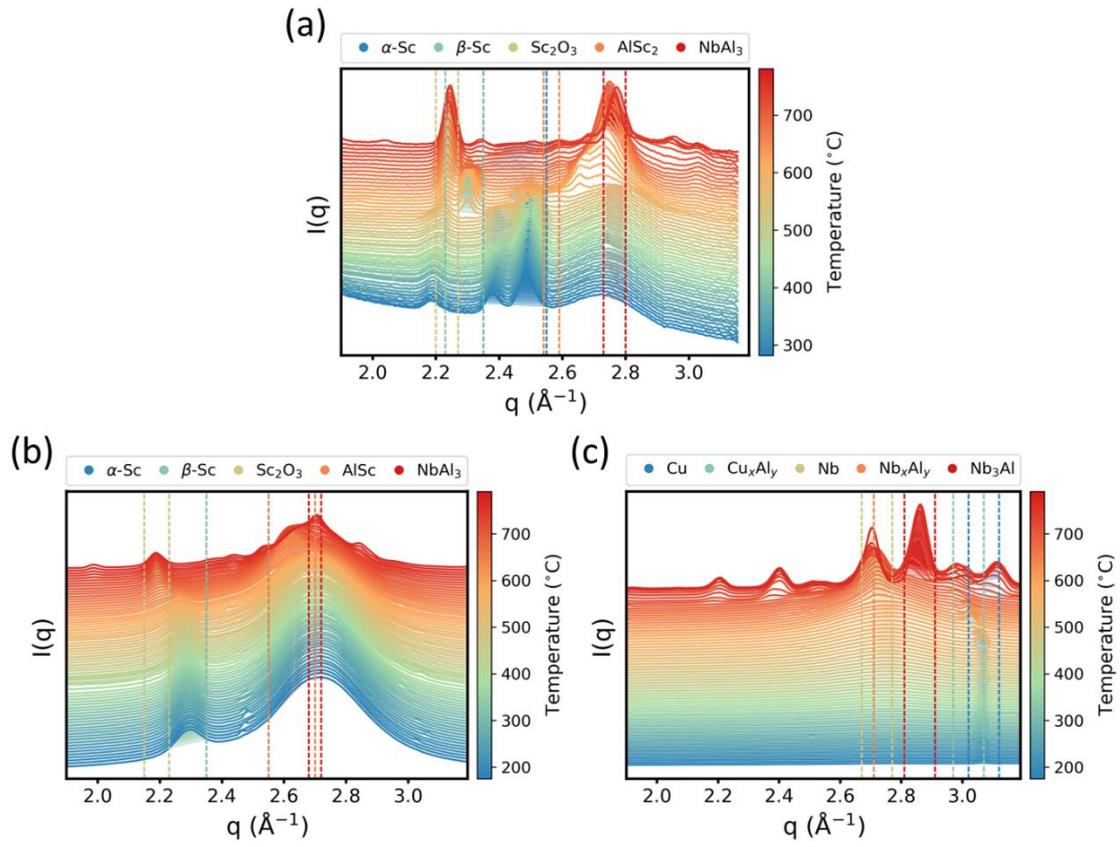

**Figure S 3.** Temperature-dependent GIWAXS patterns of stripe samples: (a) Nb-Al/α-Sc, (b) Nb-Al/β-Sc, and (c) Nb-Al/Cu thin films.

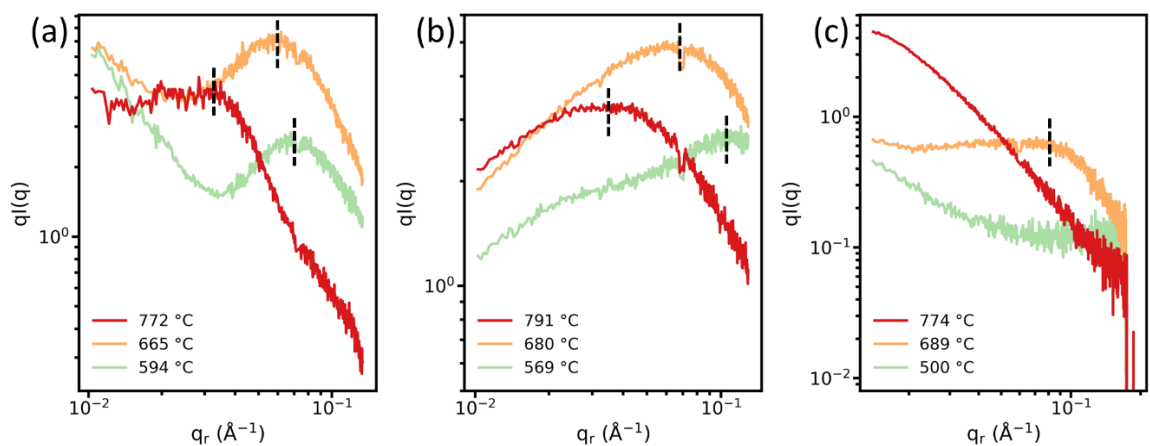

**Figure S 4.** Evolution of the characteristic length across stripe samples of (a) Nb-Al/α-Sc, (b) Nb-Al/β-Sc and (c) Nb-Al/Cu thin film at incident angle of 0.25 degree, from low to high temperatures via grazing-incidence small-angle (GISAXS) X-ray scattering.

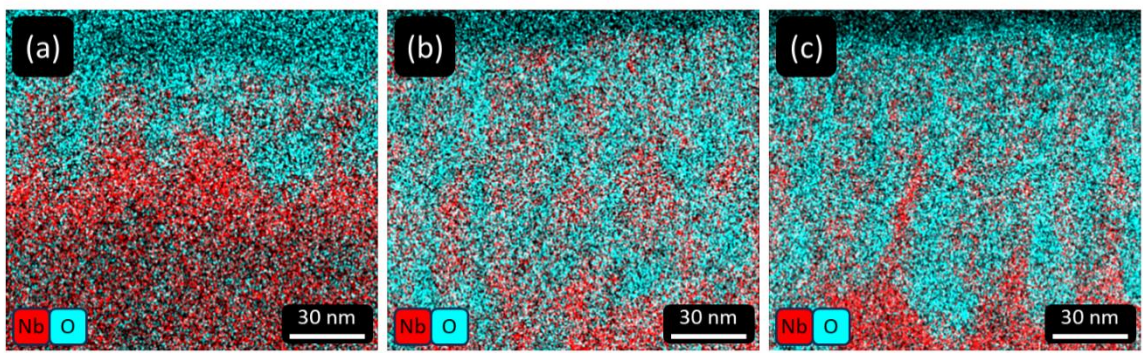

**Figure S 5.** STEM characterization using HAADF EDX for the Nb and O elements in (a) Nb-Al/α-Sc, (b) Nb-Al/β-Sc, and (c) Nb-Al/Cu thin films at 773°C, 791°C, and 732°C, respectively, corresponding to the **Figure 6**.